\documentclass[useAMS,usenatbib]{mn2e}
\usepackage{graphicx}
\usepackage{graphics}
\usepackage{epstopdf}
\usepackage{boxedminipage}
\usepackage{float}
\usepackage{url}
\usepackage{bm}
\usepackage{epsf}
\usepackage{url}

\title[Unfolding the Hierarchy of Voids]{Unfolding the Hierarchy of Voids}

\author[Aragon-Calvo M.A. et al.]{M.A. Aragon-Calvo$^{1}$\thanks{E-mail:miguel@pha.jhu.edu} ,
van de Weygaert R.$^{2}$, Araya-Melo P. $^{3}$, Platen E. $^{2}$,  Szalay A. S.$^{1}$\\
$^{1}$The Johns Hopkins University, 3400 Charles St., Baltimore, MD, USA\\
$^{2}$Kapteyn Institute, University of Groningen, the Netherlands\\
$^3$Jacobs University Bremen, Campus Ring 1, 28759 Bremen, Germany}
\begin{document}

\date{Submitted to MNRAS letters}
\pagerange{\pageref{firstpage}--\pageref{lastpage}} \pubyear{2010}
\maketitle
\begin{abstract}
We present a framework for the hierarchical identification and characterization of voids based on 
the Watershed Void Finder. The Hierarchical Void Finder is based
on a generalization of the scale space of a density field invoked in order to trace the hierarchical nature and structure of
cosmological voids. At each level of the hierarchy, the watershed transform is used to identify the voids at that particular
scale. By identifying the overlapping regions between watershed basins in adjacent levels, the hierarchical void tree is constructed. Applications on a hierarchical 
Voronoi model and on a set of cosmological simulations illustrate its potential. 
\end{abstract}

\section{Introduction}
The large scale distribution of matter observed in galaxy surveys and N-body computer simulations features a complex 
system of cell-like empty regions defined by a dense network of clusters, filaments and walls 
\citep{Kirshner81,Colless03,Huchra05,Gott05}. The Cosmic Web is the result of the 
tidally induced anisotropic nature of the gravitational collapse of density perturbations \citep{Zeldovich70,Bond96}. 

Within this context, voids are the low density depressions from which matter is continuously draining 
\citep{Icke84}. Forming a key component of the {\it Cosmic Web}, voids emerge out of the density troughs 
in the primordial Gaussian field of density fluctuations \citep[see][for a recent review]{Weyplaten09}. 
As a result of their underdensity, voids represent a region of weaker gravity, resulting in an 
effective repulsive peculiar gravitational influence. Initially underdense regions expand faster 
than the Hubble flow and while they expand, matter is squeezed in between them, resulting in  
void boundaries consisting of sheets and filaments. 

\subsection{A hierarchy of voids}
In addition to its anisotropic nature, the Cosmic Web is also characterized by an 
evident \textit{hierarchical} structure. As a result of the multiscale nature of the 
primordial perturbations, structure builds up via small scale objects into ever larger 
structures. High resolution N-body experiments \citep[][]{Springel05}
display a complex and tenuous network of substructures within the interior of voids, 
resembling the prominent Cosmic Web delineated by large haloes. The 
relation between different levels in the hierarchy of the Cosmic Web can be defined 
by the voids as, at any given level of the hierarchy, they are the cells within 
which we observe the weblike infrastructure at the next level.  

Because of their relatively simple structure and evolution, we may better understand the 
gradual hierarchical buildup of the Cosmic Web on the basis of its void population. 
Two processes dictate the evolution of voids: their {\it merging} into ever larger voids as well 
as the {\it collapse} and disappearance of small ones embedded in overdense regions. When adjacent 
voids meet up and merge, the matter in between is squeezed in thin walls and filaments, which 
subsequently drain towards the outer boundary of the voids \citep{Dubinski93}. By identifying and 
assigning critical density values to the two evolutionary void processes of merging and collapse, 
\cite{Shethwey04} managed to describe this hierarchical evolution of the void population in terms 
of a two-barrier {\it excursion set} formulation \citep{Bond91}. The context of this unfolding 
void hierarchy within the Cosmic Web can be clearly understood within the Lagrangian adhesion 
description \citep{Sahni94}. 

\subsection{Reconstructing the Hierarchy of Voids}
In this study we describe our formalism for explicitly analyzing the hierarchy of 
voids in the cosmic matter or galaxy distribution. Based on the watershed transform 
\citep[see e.g.][]{Beucher82,Platen07}, it combines the Watershed Void 
Finder (WVF) with a formalism to establish the hierarchical structure and relationship 
of the detected voids. 

The grid-based WVF method introduced by \citet{Platen07} is able to detect voids without 
restriction on their size and shape. A related Voronoi tessellation-based implementation is
the ZOBOV void finder \citep{Mark08}. Both methods are based on the idea following 
the slope lines connecting a given point in space to the local minima of the valley containing that
point. More details on the performance of a variety of other void finders can be found 
in \cite{Colberg08} \citep[also see][]{Lavaux10}. 

In section~\ref{sec:voidhier} we describe the basis of the hierarchical void tree formalism. 
The details of the technique are outlined in sect.~\ref{sec:reconstruct}. 
We then present an illustrative test of its performance on a heuristic hierarchical Voronoi model 
in section~\ref{sec:voronoi}. Its cosmological potential is outlined in sect.~\ref{sec:application}, 
followed by a short discussion in sect.~\ref{sec:conclusions}. 

\section{the Hierarchical Void Tree}
\label{sec:voidhier}
In the void hierarchy framework we identify voids independently at all levels of the hierarchical space, and establish the 
cross-scale relations between voids at different levels. For establishing the multiscale and nesting properties of the 
void network, we follow the natural path of multiscale techniques \citep{Aragon07}. Within this context, we evaluate 
the structure of a scalar field in $N$ dimensions in an $N+1$-dimensional hierarchical-space of the original field 
where the extra dimension represents a scale usually defined by a smoothing function \citep{Iijima62,Witkin83}. 

Subsequently voids between adjacent levels in the hierarchy are linked as a function of well-defined characteristics. 
A given {\it parent} void at the hierarchy level $i$ is defined by smaller {\it children} voids at the next level, $i+1$, 
in the hierarchy. We assign parent-child relations between voids in adjacent levels of the hierarchy by identifying 
overlapping volumes between the voids. A given child usually shares volume with several parent voids higher in 
the hierarchy. We enforce a {\it non-loop} property in the hierarchical tree by assigning each child void exclusively 
to the one parent void to whom the child contributes most of its volume. This constraint assures that all children 
voids have only one single parent in the \textit{void tree hierarchy}. 

\section{Reconstructing the Void Hierarchy}
\label{sec:reconstruct}
Having established the general scheme for the void hierarchy tree, we need to detail its key ingredients. The 
first issue is that of the definition of the scale-space from which we extract the void hierarchy. The most 
essential element is the void identification at each level, which is based on the watershed segmentation of 
the scalar density - or related - fields.

\subsection{Scale and Hierarchical Spaces}
Proper scale spaces must have the following set of properties: 1) linearity, 2) spatial shift invariance, 3) isotropy and 4) causality. 
The Gaussian filter addresses each of these constraints \citep{Florack93}. However, while the Gaussian 
function is an optimal scale-space operator, it is not necessarily the only - or the best suited - option for the 
study of the hierarchical character of the Cosmic Web. 

The spatial filtering approach assumes that the levels of the hierarchy are defined purely and \textit{only} on the basis of 
their corresponding spatial scale. However, it would be better if our definition of a  a characteristic 
hierarchy level was based on the nature of the complex physical processes that give rise to the dark matter and galaxy 
distribution. Intrinsic hierarchical properties of the Cosmic Web such as halo mass functions, galaxy luminosities, 
galaxy morphology, etc. are suggestive examples. In the following, the term \textit{Hierarchical Spaces} is used to indicate 
a broader class of spaces defined by one or more specific properties which are manifestations of the hierarchical nature 
of the Cosmic Web. This means they do not necessarily satisfy the requirements of a proper scale-space. 


In the case of N-body simulations we have access to the full evolution of the Cosmic Web. This allows us to 
control the relation between scales in the primordial density field. By using the information from the 
power spectrum, we can select those scales in the initial conditions which will grow and evolve faster or, 
alternatively, those that will not evolve at all. 
The most straightforward example would be the definition of a linear-regime smoothing procedure that will allow 
large-scale linear fluctuations to grow while small-scale linear fluctuations will be suppressed. This filter will 
act on the linear-regime matter distribution where all Fourier modes are independent and grow independently,  
and allow us to target specific hierarchy levels for further evolution towards collapse, ultimately producing the 
present-time structures. This low-pass filtered density field will evolve into a Universe with all the 
large scale structures in place, with their shapes moulded by anisotropic gravitational collapse, but lacking 
the small-scale details. 

This approach is fundamentally different from the usual a posteriori smoothing operation, in that 
it avoids the nonlinear effects resulting from cross talk between Fourier modes. It has the 
advantage of transparently exposing the hierarchy of structures imprinted in the initial density field.

\subsection{Watershed Segmentation}
The watershed transform segments an image into regions following its intrinsic substructure 
\citep[see][for a detailed description of the method]{Platen07}. The word {\it watershed} finds its origin in the 
analogy of the procedure with that of a landscape being flooded by a rising level of water: as the water-level 
rises, the watershed basins around the minima will ultimately meet at the ridges defined by {\it saddle-points} and 
{\it maxima} in the density field. The final result of the completely immersed landscape is a division of the 
landscape into individual cells, separated by {\it ridge dams}. The cosmological analogy to the landscape 
is suggestive: the basins represent the underdense void regions, while their boundaries of sheets and ridges 
form the network of walls, filaments and clusters that defines the Cosmic Web \citep{Aragon08}. 

\subsubsection{Oversegmentation}
One of the practical complications of watershed segmentation is its sensitivity to any structure, whether 
it is real or an artefact. As a result, it easily partitions a given region into several smaller sub-regions. 
This ``oversegmentation" is commonly assumed to be the result of ``noisy'' structures superimposed on top of 
the more prominent -- and usually ``real'' -- features. In reality, the oversegmentation is not only set 
by the noise level of the image, but also by the presence of intrinsic and significant substructure in the 
field. 

The limitations of the watershed transform due to oversegmentation can be alleviated by the use of
hierarchical techniques such as the hierarchical watershed \citep{Olsen96,Olsen97,Gauch99}. 
In this approach the watershed transform is computed on the image after smoothing at several scales or 
thresholding at several intensity levels. The large scale images will delineate large regions while smoothing 
their boundaries. The small scale images will reveal the small features in the image, as well as the noisy 
structures, while keeping the original boundaries. In a final step, the scale images are merged following 
a specific prescription. Often this involves the merging of small regions contained within a common 
parent region.

In most watershed based hierarchical reconstruction schemes, the small scale images in the hierarchy are 
merely considered as an intermediate step in the reconstruction of the features of interest. 
The oversegmentation is considered an undesirable effect due to the noise in the image. Here we will use a 
different approach. Assuming we may ignore the noise-induced oversegmentation \citep[see][]{Platen07}, 
we focus exclusively on the oversegmentation due to intrinsic structures. Instead of using only the 
largest scale of the hierarchy, we therefore will consider all scales simultaneously.

\subsubsection{Hierarchical Watershed}

We perform the void merging across adjacent levels in the hierarchy by computing only the flooding procedure on the watershed,  
i.e. without identifying the watershed boundaries. This procedure segments the density field into watershed basins but 
does not explicitly provide the boundaries between adjacent watershed regions. This ``incomplete watershed" focuses 
only on the space partitioning aspect of the watershed transform. This makes it straightforward to merge voxels between 
children voids on the basis of this incomplete watershed. This leads directly to the complete hierarchical void tree. 
After the merging procedure for completeness we compute the full watershed transform (i.e. watershed basins and boundaries)
by performing a local flooding watershed transform restricted to the boundary voxels as described in \citet{Aragon08}.

Once the void hierarchy is stored in a tree structure it is straightforward to define functions to transverse the tree 
and extract useful information of the properties of the voids, their connectivity and their hierarchical relations. 

\section{Test: Hierarchical Voronoi Models}
\label{sec:voronoi}
We tested our method with a hierarchical implementation of a Voronoi clustering model of the Cosmic Web 
\citep{Okabe2000,Weyicke89}. This model shares similar spatial and hierarchical properties as the observed distribution of matter
while making it possible to objectively compare the recovered void hierarchy with the original one.
Hierarchical Voronoi models used have the two main properties we seek to study: 
1) a clear multiscale nature and 2) a hierarchy of nested structures.

\subsection{Implementation}
The hierarchical Voronoi model is constructed as follows: the top level of the void hierarchy is generated from a 
set of sparsely sampled points which define a periodic Voronoi tessellation. 
Inside each Voronoi cell we define a new set of points and compute the
Voronoi tessellation \textit{locally} on the points inside the cell. This local Voronoi cell is non-periodic 
and has its parent Voronoi edges as boundaries. This procedure can be repeated
iteratively until the desired number of nested levels in the hierarchy is reached.
We regularize the size and shape of the Voronoi cells by performing a Voronoi Centroid regularization 
on the seed points.  By coupling the scalar  and hierarchical aspects of the image we can study 
it via the canonical Gaussian scale-space.

\begin{figure*}
  \centering
  \vskip -0.25truecm
  \includegraphics[width=0.98\textwidth]{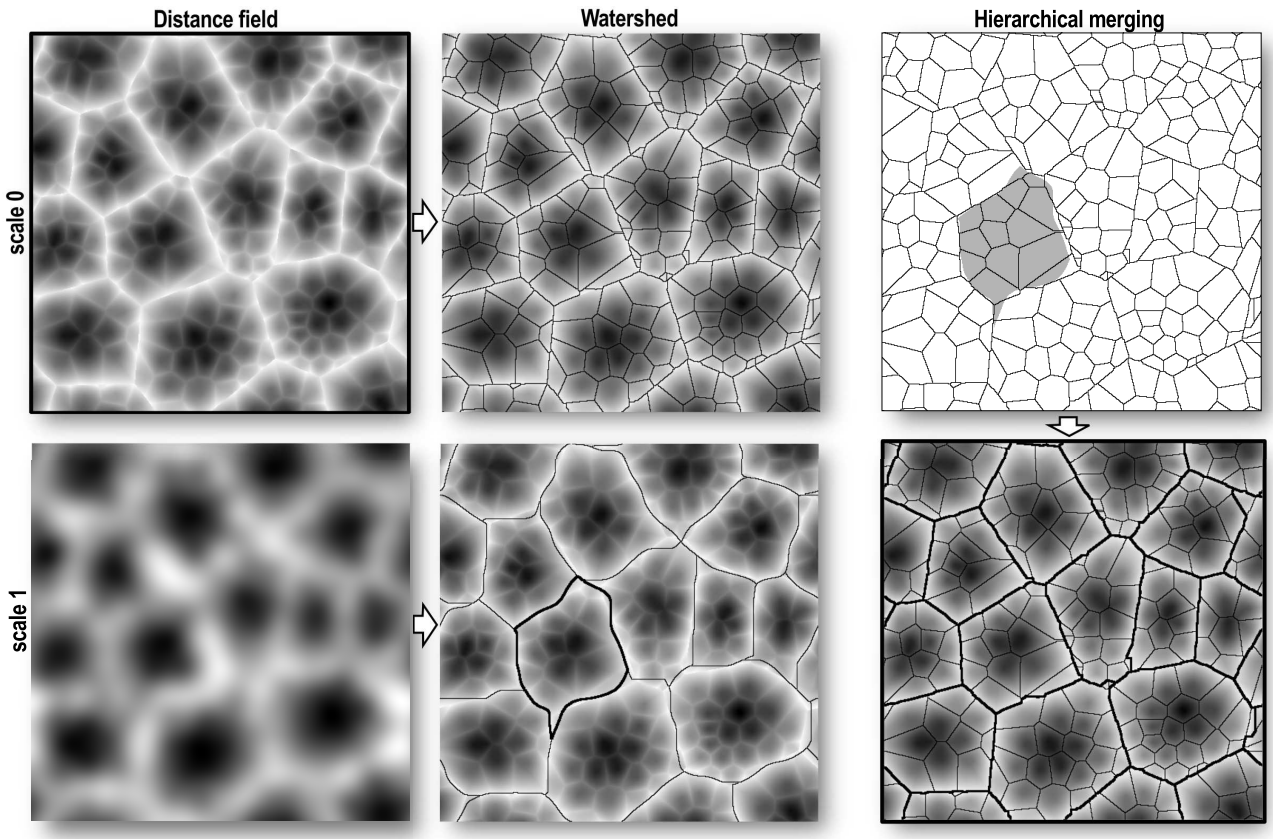}
    \caption{Hierarchical reconstruction of voids in a hierarchical Voronoi model.
      The original distance field is shown in the top left panel (scale $0$) and its smoothed version (scale 1) in the lower left panel.
      the center panels show their corresponding watershed transform.  An individual  void is depicted at the largest scale in the
      center low panel. The hierarchical merging of the void with its children sub-voids is shown in the top right panel.
      The final hierarchical reconstruction is shown in the bottom right panel. The original shape of the large voids is 
      reconstructed as well as their inner hierarchy of substructures.}
  \label{fig:Void_merging_overlap}
\end{figure*}

\subsection{the Voronoi Test}
From the hierarchical Voronoi model we compute the \textit{normalized distance field} for each point
in a regular grid (see \citet{Aragon08}). This field is defined as the ratio between the distance to the \textit{closest} and \textit{second closest} 
Voronoi seed. It yields a distance field with values of 1 at the cell boundaries and
decreasing towards the center of the cell. We do this for each level in the hierarchy. Finally, all levels 
are integrated into a single distance field as
$\mathcal{I}(\vec x)$ as $\mathcal{I}(\vec x) = \sum^{n}_{i=0} (\mathcal{I}^i(\vec x)/(i+1)^2)$, 
where $n$ is the number of levels in the hierarchy. 
This scale integration scheme is similar to the one used in other synthetic image generation algorithms such as Perlin noise \citep{Perlin85}.
By tuning the denominator of the above equation it is possible to define different intensity scaling relations between
levels in the hierarchy. In our case the most prominent features in the image will be the largest voids.

Next, we construct the Gaussian scale space of the image and identify voids at each scale independently. Since our image was constructed 
with two characteristic scales it makes no sense to use more than two smoothing scales. The scale-space then consists of two scales, one 
with no smoothing and one with a width between the size of the small and large Voronoi cells.  

The hierarchical merging of voids is illustrated in figure \ref{fig:Void_merging_overlap}. The left top and bottom panels show the original
field and its smoothed version respectively. 
The center panels show the corresponding watershed transform. Note that the smoothed
field produces a distorted watershed transform. Both the general shape and the boundaries of the voids are affected by the smoothing procedure.
On the other hand, the watershed transform of the original field reproduces the original boundaries between voids but it does not differentiate between levels 
in the hierarchy. The hierarchical merging of voids is shown in the top right panel. One individual void is highlighted 
in order to illustrate the void merging procedure. The parent void's area overlaps with several children and one can see that 
the children that are mostly covered by the parent void are the ones originally inside it. The final result of the merging 
procedure is shown in the  bottom right panel where we emphasize the large voids in the top level of the hierarchy (thick lines) 
containing smaller voids at the bottom of the hierarchy (thin lines).

The hierarchical reconstruction of the voids has two important advantages over the single-scale watershed void finder: 1) It is not affected by smoothing
procedures and 2) it explicitly gives the inner substructure of the voids. The reconstructed hierarchical voids contain both their original shape and their original
level of substructure.

\section{Cosmological Application}
\label{sec:application}
We applied our algorithm to three cosmological simulations that are 
variants of the cold dark matter scenario.
The simulations cover the three possible geometries of the Universe: flat, 
open and closed, with cosmological parameter of ($\Omega_{m}$, 
$\Omega_{\Lambda}$) = (0.3,0.7) for the the flat $\Lambda$CDM model, (0.1,0.7) 
for the open $\Lambda$CDM and (0.5,07) for the closed $\Lambda$CDM Universe.
Each simulation consists of $256^{3}$ dark matter particles in a 200$h^{-1}$Mpc box. 
All simulations share the same Hubble parameter, $h=0.7$  and $\sigma_{8}=0.8$
\citep[see][for a detailed description]{Araya08}.

\begin{figure*}
  \centering
  \vskip -0.25truecm
  \includegraphics[width=0.31\textwidth]{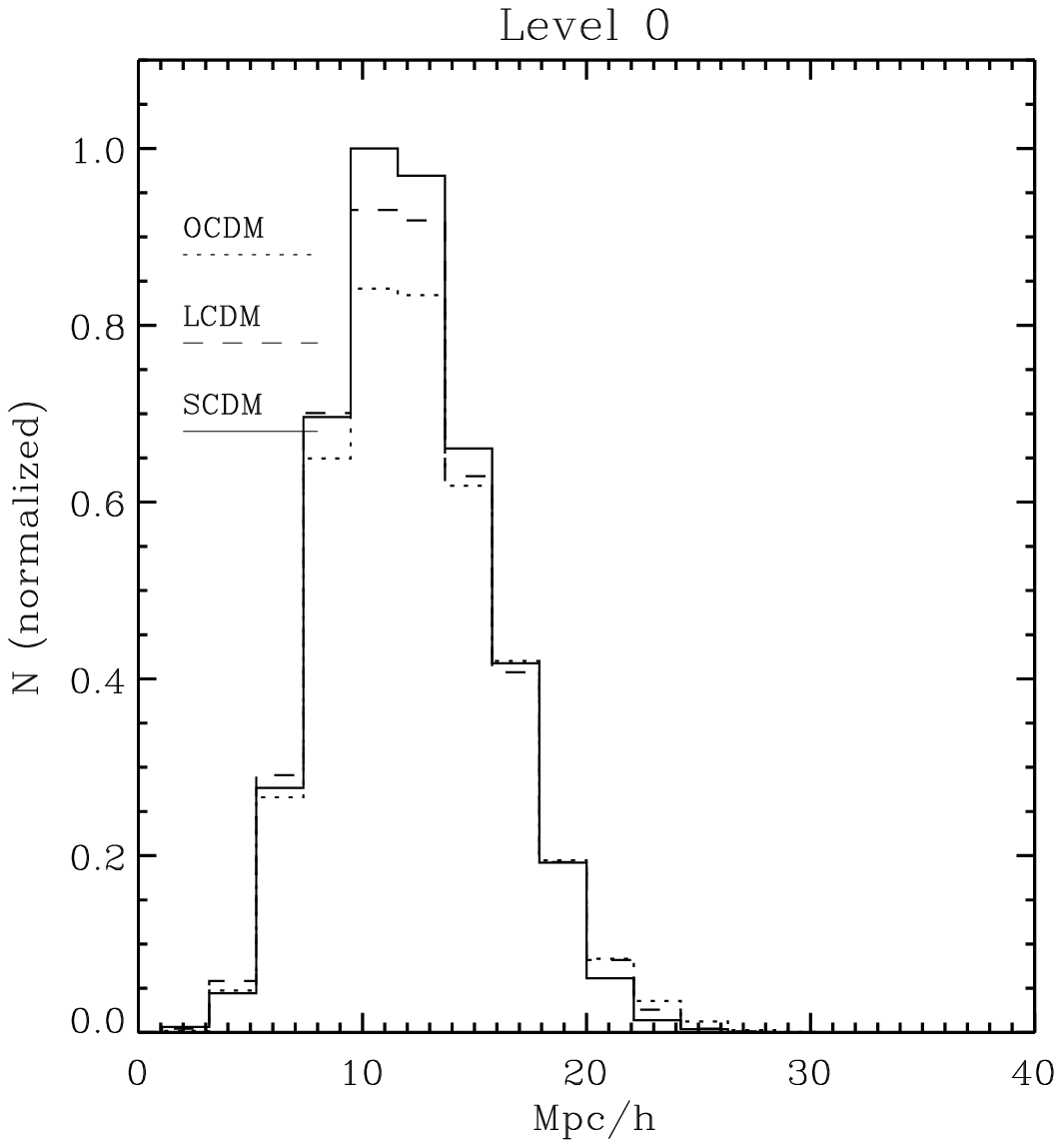}
  \includegraphics[width=0.31\textwidth]{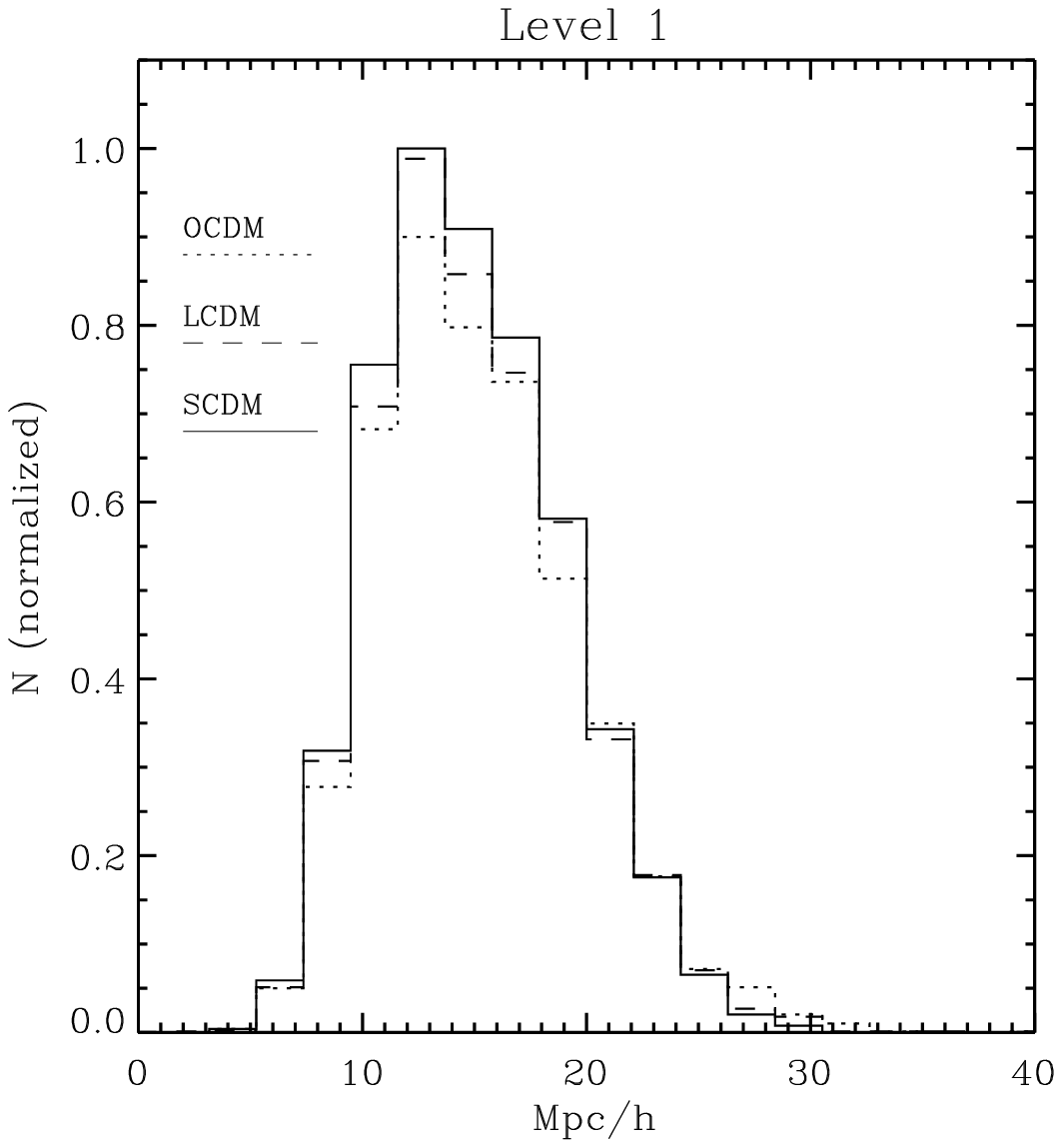}
  \includegraphics[width=0.31\textwidth]{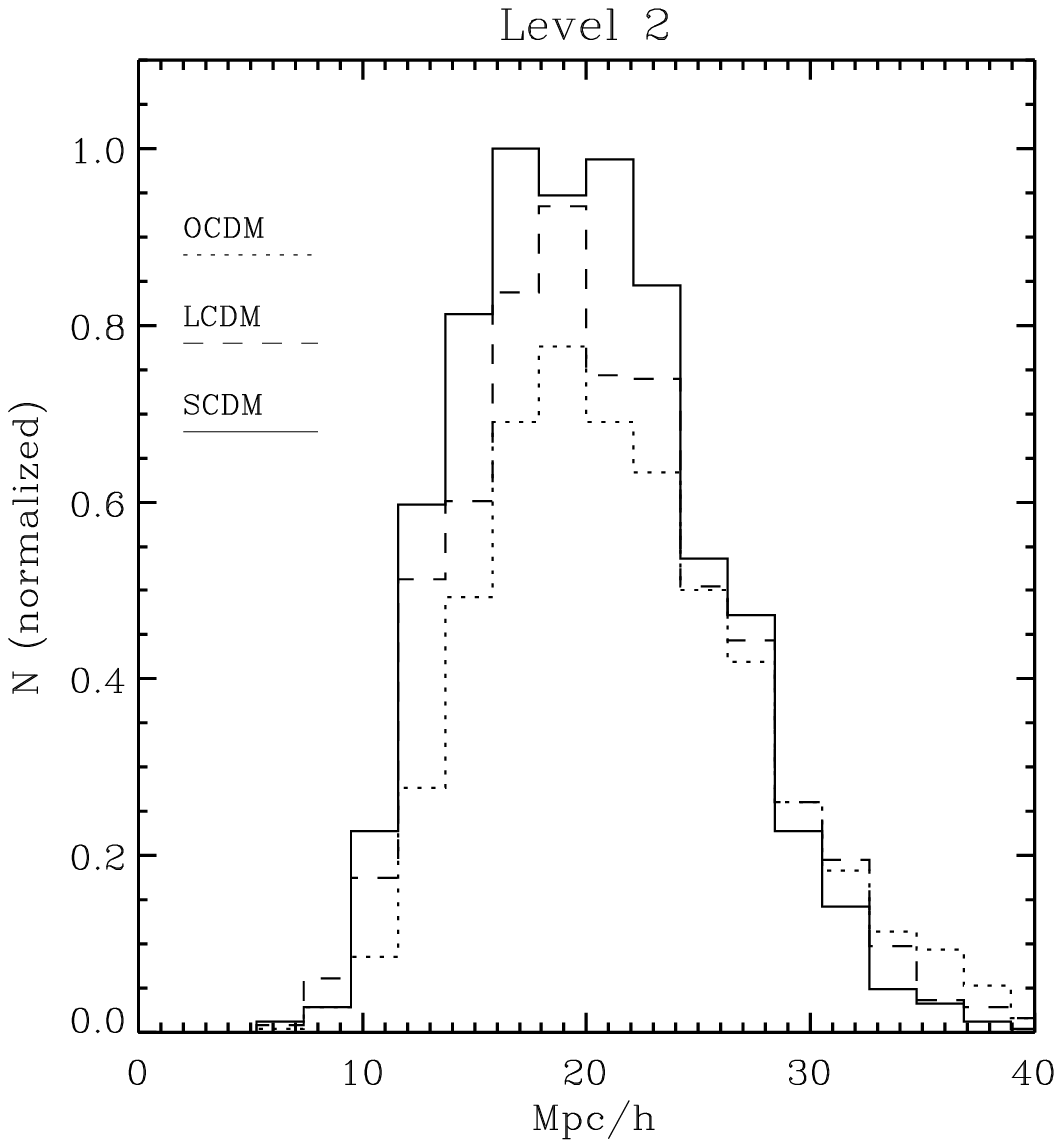}
    \caption[1]{ \small Distribution of void sizes for three cosmologies: SCDM (solid) $\Lambda$CDM (dashed) and OCDM (dotted) computed
    at three different levels of the hierarchy going from the (left) bottom of the hierarchy (smallest scale) to (right) the top of the hierarchy. }
  \label{fig:trees_compare_size_distribution}
\end{figure*}

We perform the linear-regime smoothing procedure by generating lower-resolution versions of $128^3$ and
 $64^3$ particles from the same  initial conditions. The $64^3$ resolution corresponds to a cut-off scale 
of $\sim 3h^{-1}$Mpc, enough to trace voids without significant substructure. 
We followed the evolution of the box from $z=49$ until the present  time, $z=0$, using the GADGET-2 N-body code \citep{Springel05}. 
From the final particle distribution we compute the density field inside a cubic grid of 512 voxels per 
dimension using a recent implementation of the DTFE method \citep{Schaapwey00, Weyschaap09}.

The size distribution of voids in different cosmologies and levels of the hierarchy are shown in figure 
\ref{fig:trees_compare_size_distribution}. The mean void sizes of voids in all cosmologies are 11,13 and 18 Mpc/$h$ for 
levels 0,1 and 2 respectively. While the voids at the top of the hierarchy 
(level 2) are clearly the largest, the mean size and distribution of voids in levels 0 and 1 are very similar. 
All distributions in the three cosmologies have similar peaks at a given level in the hierarchy. However, there are differences in the overall shape of the distributions. Compared to the LCDM and SCDM, the OCDM universe has a higher tail towards large voids. The fact that the
order OCDM-LCDM-SCDM is observed in all the distributions gives us a good indication of the ability of our
method to discriminate between cosmologies. 

\section{Conclusion and future work}
\label{sec:conclusions}
We introduced a framework for the identification of voids and their hierarchical 
properties.  The hierarchical nature of the void network makes our method a powerful tool for
its description and characterization. The Hierarchical Void Finder shares the advantages of 
the Watershed Void Finder, while addressing some of its limitations such as the
oversegmentation and the reconstruction of the void boundaries after strong 
smoothing of the density field.

In order to test our method we introduced a hierarchical implementation of Voronoi models. 
These heuristic models share the multiscale, hierarchical  and topological properties of the Cosmic Web.
As such the hierarchical Voronoi models represent a valuable tool for testing algorithms for LSS analysis.

We extend the idea of scale-space in order to account for non linearities and physical processes. 
We discuss a Gaussian smoothing in the initial conditions.
By applying, a Gaussian smoothing in the linear regime before there is cross-talk between Fourier modes we 
are able to cleanly expose the hierarchy of structures in the evolved non-linear matter distribution.

In a following paper we will describe the basis of the hierarchical space and explore in more detail the 
properties of the void network and its potential for constraining cosmological parameters.

\end{document}